\documentclass[10pt,aps,prd,onecolumn,showpacs,amsmath,amssymb,nofootinbib,eqsecnum,preprintnumbers,superscriptaddress]{revtex4-2}

\usepackage{mathtools}
\usepackage{xcolor}
\usepackage{graphicx}
\usepackage{tikz}
\usepackage{comment}
\usepackage[colorlinks=true,
            citecolor=red,
            linkcolor=blue,
            urlcolor=violet,
            filecolor=cyan,
            backref=false]{hyperref}
\usepackage{multirow}
\usepackage{titlesec}

\usepackage{enumerate}

\titleformat{\paragraph}[block]{\filcenter}{}{0pt}{}

\newcommand\feq{\mathrel{\phantom{=}}}

\begin{document}


\title{
Uniqueness of Galilean and Carrollian limits of gravitational theories and application to higher derivative gravity
}

\author{Poula Tadros}
\email{poula.tadros@matfyz.cuni.cz}
\affiliation{Institute of Theoretical Physics, Faculty of Mathematics and Physics, Charles University,
V Hole\v{s}ovi\v{c}k\'ach 2, Prague 180 00, Czech Republic}

\author{Ivan Kol\'a\v{r}}
\email{ivan.kolar@matfyz.cuni.cz}
\affiliation{Institute of Theoretical Physics, Faculty of Mathematics and Physics, Charles University,
V Hole\v{s}ovi\v{c}k\'ach 2, Prague 180 00, Czech Republic}

\date{\today}

\begin{abstract}
    We show that the seemingly different methods used to derive non-Lorentzian (Galilean and Carrollian) gravitational theories from Lorentzian ones are equivalent. Specifically, the pre-nonrelativistic and the pre-ultralocal parametrizations can be constructed from the gauging of the Galilei and Carroll algebras, respectively. Also, the pre-ultralocal approach of taking the Carrollian limit is equivalent to performing the ADM decomposition and then setting the signature of the Lorentzian manifold to zero. We use this uniqueness to write a generic expansion for the curvature tensors and construct Galilean and Carrollian limits of all metric theories of gravity of finite order ranging from the $f(R)$ gravity to a completely generic higher derivative theory, the $f(g_{\mu\nu},R_{\mu\nu\sigma \rho},\nabla_{\mu})$ gravity. We present an algorithm for calculation of the $n$-th order of the Galilean and Carrollian expansions that transforms this problem into a constrained optimization problem. We also derive the condition under which a gravitational theory becomes a modification of general relativity in both limits simultaneously.
\end{abstract}

\maketitle

\section{Introduction}
The \textit{Galilean limit} of general relativity (GR) was first computed in \cite{Dautcourt:1996pm} as an extension of \textit{Newton-Cartan (NC) theory} by taking the limit as the speed of light goes to infinity, $c\to\infty$. Later development were made in \cite{Tichy:2011te,VandenBleeken:2017rij} showing that the leading order (LO) in the Galilean limit gives the Newton-Cartan structure, but the next-to-leading order (NLO) gives the \textit{type II torsional Newton-Cartan (TNC) geometry} \cite{Hansen:2019pkl}. With the growing interest in non-Lorentzian geometry in the recent years, there have been many papers studying this limit \cite{Bergshoeff:2020fiz,Cariglia:2018hyr,Bergshoeff:2021bmc,Concha:2022jdc,Bergshoeff:2015uaa}. The methods of taking the Galilean limit of a relativistic theory are
\begin{enumerate}[i)]
    \item By using the \textit{pre-nonrelativistic (PNR) parametrization} of the metric which was introduced in \cite{Hansen:2020pqs} and further used in \cite{Guerrieri:2020vhp}; for a review see \cite{10.3389/fphy.2023.1116888}.
    \item By performing the ADM decomposition, then taking the limit of the norm of the normal vector to infinity. We call it the \textit{``infinite scaling'' (IS)} method.
    \item By gauging the Bargmann algebra \cite{Andringa:2010it}. This is motivated by the fact that the Galilean algebra, which on gauging gives NC geometry, can be constructed by the nonrelativistic In\"on\"u-Wigner contraction of the Poincar\'e group. On extending it to the Bargmann algebra, the gauging procedure gives TNC geometry. This method, known as \textit{Galilean algebra gauging (GAG)}, was applied to study the Galilean limit of GR in \cite{Bergshoeff:2017btm}.
\end{enumerate}
A different method was proposed in \cite{Banerjee:2022eaj} where covariant and contravariant tensors are scaled differently (by $c^{-1}$ and $c$ respectively), although this method makes it easier to take the Galilean limits of Maxwell's theory to the LO and NLO, it is equivalent to the PNR approach since lowering and raising indices is done by the metric with a factor of $c^2$ and its inverse with a factor of $c^{-2}$, and the two approaches give the same Galilean theories. The localization of Galilean symmetries are done in \cite{Banerjee:2014pya,Banerjee:2016laq,Banerjee:2014nja}. Since then, the Galilean limit has been used in condensed matter \cite{PhysRevD.97.125006,Geracie:2016bkg}, fluid mechanics \cite{PhysRevD.93.065007,BANERJEE20191} and string theory \cite{Oling:2022fft,PhysRevLett.128.021602,Hartong:2022dsx} where the NC and TNC geometries are promoted to their stringy versions. 

The \textit{Carrollian limit} is the opposite limit to the Galilean one. It was first considered as the ultralocal In\"on\"u-Wigner contraction of the Poincar\'e group in the mid 60s by Levy-Leblond \cite{Levy-Leblond}, and independently by Sen Gupta \cite{Gupta} where they considered the limit as the speed of light approaches zero, ${c\to0}$. However, there was no physical interpretation nor application of this limit for more than 40 years. It was only considered in physics papers in the 2000s in various but yet limited cases in conformal field theories and ultrarelativistic fluids. The Carrollian limit became more popular among physicists when a direct connection with physics near black hole horizons was established in 2019 in \cite{Donnay:2019jiz}. It allowed for defining physical quantities on black hole horizons, while preventing the divergences present in the membrane paradigm \cite{Price:1986yy,1986bhmp.book.....T,Damour:1978cg}.

As of today, the Carrollian limit has been applied in various areas of theoretical physics. Specifically, the Carrollian physics and Carrollian structures were analyzed in the context of representations of the Carroll group, i.e, the Carroll particles \cite{Zhang:2023jbi,Bergshoeff_2014,Marsot:2021tvq,Marsot:2022imf}. In condensed matter physics \cite{Bagchi:2022eui,Kubakaddi_2021,Kononov_2021}, the Carrollian limit was used to study fractons and its continuum field limit \cite{Baig:2023yaz,Kasikci:2023tvs}. It also found various applications in field theory \cite{PhysRevD.106.085004,Chen:2023pqf,Bergshoeff:2022eog,Henneaux:2021yzg} and conformal field theory \cite{Bagchi:2019xfx,Bagchi:2019clu,Bagchi:2021gai,PhysRevD.103.105001}, such as the holography. The Carrollian limit of fluids \cite{Bagchi:2023ysc,Ciambelli:2018wre,Ciambelli:2018xat,campoleoni2019two,Ciambelli:2020eba,10.21468/SciPostPhys.9.2.018} was found particularly interesting as it allowed non trivial motion even in the LO. Some cosmological applications of the limit were also considered \cite{deBoer:2021jej,Bonga:2020fhx}. The Carrollian limit was further used in string theory \cite{PhysRevLett.123.111601,Bagchi:2021ban,Cardona:2016ytk}, gravity \cite{Perez:2021abf,Perez:2022jpr,Hartong:2015xda,Figueroa-OFarrill:2021sxz,Hansen:2021fxi,Gomis:2020wxp,Bergshoeff:2022qkx,Guerrieri:2021cdz,Hansen:2020wqw}, black holes \cite{Donnay:2019jiz,Grumiller:2019tyl,Redondo-Yuste:2022czg,Marsot:2022imf,Anabalon:2021wjy}; particularly to analyze dynamics of particles near black-hole horizons \cite{Gray:2022svz,Bicak:2023vxs,Marsot:2022qkx}. Recently, the Carrollian limit gave rise to a new theory of holography on null boundaries of asymptotically flat spacetimes \cite{Herfray:2021qmp,Chandrasekaran:2021hxc,Bagchi:2019clu,Ciambelli:2019lap}.

While calculating the Carrollian limit of Maxwell's theory, it was shown that there are two nonequivalent Carrollian limits for any Lorentzian theory. The first one keeps the electric part in Maxwell's theory and so was named the \textit{``electric limit''}, and the other keeps the magnetic part and so was called the \textit{``magnetic limit''}. In gravitational theories, the electric limit is the LO in the Carrollian expansion of the Lagrangian and the magnetic limit is the NLO in that expansion. There are multiple ways to take the (electric and magnetic) Carrollian limit of a Lorentzian gravitational theory:
\begin{enumerate}[i)]
    \item By parametrizing the Lagrangian using the \textit{pre-ultralocal (PUL) parametrization}, then taking $c \to 0$  \cite{Hansen:2021fxi,deBoer:2023fnj}.
    \item By performing the ADM decomposition for the Lagrangian and set the signature of the metric to zero \cite{Henneaux:1979vn,Henneaux:2021yzg}, which we refer to as the \textit{``zero signature'' (ZS)} approach.
    \item By gauging the algebra of the theory and performing the ultralocal In\"on\"u-Wigner contraction \cite{Bergshoeff:2017btm}, i.e., \textit{Carroll algebra gauging (CAG)} approach.
\end{enumerate}
Another possibility is by rescaling certain terms and taking the appropriate limit to get the desired theory \cite{Niedermaier:2023hgk}. This method was implemented for multigravity theories in \cite{Ekiz:2022wbi}. Since rescaling terms will give a subset of theories given by the PUL approach, we consider it at a special case. The Carrollian limit of GR was also constructed using the Kol-Smolkin (KS) decomposition (which is dual to the ADM decomposition), expanding the resulting quantities in powers of $c^{-1}$, then taking the limit ${c \to 0}$ \cite{Elbistan:2022plu}. This method gives the same results as the ZS approach since the two decompositions are dual, and the limit ${c \to 0}$ gives the same result as setting the signature to zero as we will show.

In this paper, we show that the PNR parametrization can be constructed from GAG. Similarly, the PUL parametrization can be constructed from CAG. We also show that the PUL and ZS approaches of taking the Carrollian limit are completely equivalent (and, similarly, PNR and IS). Throughout this paper, we consider the ADM decomposition with an extra real parameter $\epsilon$ which can be interpreted as a signature if it is $1$, $-1$, or $0$; the zero signature case is a Carrollian manifold \cite{Henneaux:1979vn}. Once we establish the uniqueness of the Galilean and Carrollian limits (i.e., the equivalence of above approaches), we construct an algorithm for computing the $n$-th order of the Galilean, and the Carrollian expansions of a completely generic higher derivative gravity (HDG). The problem is transformed into computationally easier optimization problems which upon solving gives the desired order. 

The paper is organised as follows:
\begin{itemize}
    \item In Sec.~\ref{sec2}, we briefly review the methods used to construct non-Lorentzian gravitational theories from Lorentzian ones.
    \item In Sec.~\ref{sec3}, we give a proof that the PUL and the ZS approaches are equivalent in the sense that they describe the same limit of the metric with same variables and parameters. Naturally, the same is true for PNR and IS approaches. Furthermore, we show that the PUL parametrization and the Carroll compatible connection can be constructed by gauging the Carroll algebra, i.e., the equivalence of PUL and CAG. Similarly, we also prove that we can construct the PNR parametrization and the Galilei compatible connection from the Galilei algebra gauging procedure, i.e., the equivalence of PNR and GAG.
    \item In Sec.~\ref{sec6}, while making use of the uniqueness, we write a general expansion of curvature and convert the calculation of the Galilean and Carrollian expansions of the $f(R)$ gravity, the $f(g_{\mu\nu},R_{\mu\nu\sigma \rho})$ gravity, as well as the most general HDG, i.e., the $f(g_{\mu\nu},R_{\mu\nu\sigma \rho}, \nabla_{\mu})$ gravity, into the constrained optimization problems. We also present the conditions under which a gravitational theory can be a modification of GR in both the Galilean and Carrollian regimes.
    \item In Sec.\ref{sec7}, we conclude the paper by a brief summary and discussion of our results.
\end{itemize}


\section{Review of non-Lorentzian limits}\label{sec2}
We begin with a brief review of the methods that are used in the available literature for construction of non-Lorentzian theories.

\subsection{Pre-ultralocal parametrization}
The PUL parameterization is the parametrization that is suitable for taking the ultralocal (Carrollian) limit \cite{Hansen:2021fxi}. The metric parametrization is the
given by
\begin{equation}\label{metric parametrization}
\begin{aligned}
g_{\mu \nu} &= -c^2 T_{\mu}T_{\nu}+ \Pi_{\mu \nu},& \hspace{10pt}
g^{\mu \nu} &= -\tfrac{1}{c^2} V^{\mu}V^{\nu}+ \Pi^{\mu \nu},
\end{aligned}
\end{equation}
where $T_{\mu}$ and $V^{\mu}$ is the orthonormal covector and vector while $\Pi_{\mu\nu}$ and $\Pi^{\mu\nu}$ is the induced metric and its inverse; they satisfy
\begin{equation}\label{conditions}
\begin{aligned}
T_{\mu}V^{\mu} &=-1, 
& \hspace{10pt}
-V^{\mu}T_{\nu} + \Pi^{\rho \mu} \Pi_{\rho \nu} &= \delta_{\mu}^{\nu},
& \hspace{10pt}
T_\mu \Pi^{\mu \nu} = \Pi_{\mu \nu} V^\nu&=0. 
\end{aligned}
\end{equation}
It proves useful to introduce a Carroll compatible connection by demanding that $\Pi^{\mu\nu}$ and $T_{\mu}$ are covariantly constant,
\begin{equation}\label{PUL connection}
    C^{\rho}_{\mu\nu} = -V^{\rho}\partial_{(\mu}T_{\nu)}- V^{\rho}T_{(\mu}\pounds_{\boldsymbol{V}}T_{\nu)} + \tfrac{1}{2} \Pi^{\rho \lambda}\bigg[ \partial_{\mu} \Pi_{\nu \lambda}+ \partial_{\nu} \Pi_{\lambda \mu}- \partial_{\lambda}\Pi_{\mu \nu}\bigg]- \Pi^{\rho \lambda}T_{\nu}\mathcal{K}_{\mu \lambda},
\end{equation}
where $\mathcal{K}_{\mu \lambda}$ is the extrinsic curvature defined by ${\boldsymbol{\mathcal{K}}= -\tfrac{1}{2} \pounds_{\boldsymbol{V}} \boldsymbol{\Pi}}$.  This connection is defined so that the respective covariant derivative leaves the Carrollian structure invariant (similar to the requirement that the covariant derivative of the metric is zero in the Lorentzian case). This can be gone by requiring that $\Pi_{\mu\nu}$ and $V^{\mu}$ (the defining quantities for the Carrollian structure) to be covariantly constant. Such connection can be derived from fundamental arguments in the gauging procedure of the Galilei algebra as we will see below.

Assuming analyticity, we again expand all the quantities in powers of $c^2$:
\begin{equation}
    \begin{aligned}
    \Pi^{\mu\nu} &=h^{\mu\nu}+c^2 \Phi^{\mu \nu}+ O(c^4),
& \hspace{10pt}
     \Pi_{\mu\nu} &=h_{\mu\nu}+c^2 \Phi_{\mu \nu}+ O(c^4),\\  
    V^{\mu} &=v^{\mu}+c^2 M^{\mu}+ O(c^4),
& \hspace{10pt}
    T_{\mu} &=\tau_{\mu}+c^2 N_{\mu}+ O(c^4).
    \end{aligned}
\end{equation}
This allows us to express the Lagrangian in powers of $c^2$. The LO (electric limit) will define an ultralocal theory, i.e, a theory that does not admit a non-tachyonic single-particle motion. In contrast, the theory obtained in the NLO (magnetic limit) allows some motion, and it also admits massive solutions.

\subsection{Pre-nonrelativistic parametrization}
A convenient parametrization of the metric that allows one to take the nonrelativistic (Galilean) limit is the PNR parametrization \cite{Hansen:2020wqw}. The metric parameterization is the same as \eqref{metric parametrization} with the conditions \eqref{conditions}, but a different connection that is compatible with Galilean symmetries:
\begin{equation}
    C^{\rho}_{\mu\nu} = -V^{\rho}\partial_{\mu}T_{\nu} + \tfrac{1}{2}\Pi^{\rho \sigma}\bigg(\partial_{\mu}\Pi_{\nu\sigma}  + \partial_{\nu}\Pi_{\mu\sigma} - \partial_{\sigma}\Pi_{\mu\nu}\bigg).
\end{equation}
This connection is defined so that the respective covariant derivative leaves the Galilean structure invariant. This can be gone by requiring that $\Pi^{\mu\nu}$ and $T_{\mu}$ to be covariantly constant. Such connection can be derived from fundamental arguments in the gauging procedure of the Galilei algebra as we will see below.

Now, assuming the analyticity of all the quantities, we expand $\Pi_{\mu\nu},\Pi^{\mu\nu}, V^{\mu}, T_{\mu}$ in powers of ${c^{-2}}$:
\begin{equation}
    \begin{aligned}
      \Pi_{\mu\nu} &= h_{\mu\nu} + \tfrac{1}{c^2} \Phi_{\mu\nu} + O\big(\tfrac{1}{c^4}\big), & \hspace{10pt}
      \Pi^{\mu\nu} &= h^{\mu\nu} + \tfrac{1}{c^2} \Phi^{\mu\nu} + O\big(\tfrac{1}{c^4}\big), \\
      V^{\mu} &= v^{\mu} + \tfrac{1}{c^2} m^{\mu} + O\big(\tfrac{1}{c^4}\big), & \hspace{10pt}
      T_{\mu} &= \tau_{\mu} + \tfrac{1}{c^2} m_{\mu} + O\big(\tfrac{1}{c^4}\big),
    \end{aligned}
\end{equation}
where $m_{\mu}$ is the mass current which is conserved in Galilean theories. (It also corresponds to the additional generator in the extension of Galilean algebra to Bargmann algebra in the gauging process.) The final step in the Galilean limit is to use the formulae above to expand the Lagrangian. The expansion to the LO will only involve the LO fields $h_{\mu\nu},h^{\mu\nu},v^{\mu}$ and $\tau_{\mu}$, which define NC structure, while the NLO should also include $\Phi^{\mu\nu}, \Phi_{\mu\nu}, m^{\mu}$, and $m_{\mu}$, which modifies it to type II TNC structure.

\subsection{Zero signature approach}
The ZS approach is a method of taking the ultralocal limit of a Lorentzian theory. It consists of performing the ADM decomposition, and then setting the signature of the manifold to zero. This process results in a Carrollian manifold \cite{Henneaux:1979vn}. This is equivalent of taking the limit of the norm of the normal vector to zero. The first step is to decompose the metric with a Lorentzian and a Riemannian signatures. The Lorentzian metric is decomposed as
\begin{equation}
    \begin{aligned}
        g_{\mu\nu} = -n_{\mu}n_{\nu} + h_{\mu\nu}, \hspace{10pt} &  \hspace{10pt} g^{\mu\nu} = -n^{\mu}n^{\nu} + h^{\mu\nu},
    \end{aligned}
\end{equation}
where $h_{\mu\nu}$ is the induced metric on the spatial surfaces and $n_{\mu}$ is the corresponding orthogonal 1-form with $n_{\mu}n^{\mu}=-1$.
The Riemannian counterpart is
\begin{equation}
    \begin{aligned}
        g_{\mu\nu} = n_{\mu}n_{\nu} + h_{\mu\nu}, \hspace{10pt} &  \hspace{10pt} g^{\mu\nu} = n^{\mu}n^{\nu} + h^{\mu\nu}.
    \end{aligned}
\end{equation}
Calculating the Hamiltonian of GR in both signatures yields
\begin{equation}
    H = G_{\mu\nu\sigma\rho}\pi^{\mu\nu}\pi^{\sigma \rho} - \sqrt{g}R = G_{\mu\nu\sigma\rho}\pi^{\mu\nu}\pi^{\sigma \rho} + \bar{R} - (K^2-K_{\mu\nu}K^{\mu\nu}),
\end{equation}
where $G_{\mu\nu\sigma \rho} = \tfrac{1}{2\sqrt{g}}(g_{\mu\sigma}g_{\nu\rho} + g_{\mu\rho}g_{\nu\sigma}-g_{\mu\nu}g_{\sigma\rho})$, $\bar{R}$ is the three dimensional Ricci scalar, $K_{\mu\nu}$ is the extrinsic curvature, and $K=h^{\mu\nu}K_{\mu\nu}$ in the Lorentzian case, and
\begin{equation}
     H = G_{\mu\nu\sigma\rho}\pi^{\mu\nu}\pi^{\sigma \rho} + \sqrt{g}R = G_{\mu\nu\sigma\rho}\pi^{\mu\nu}\pi^{\sigma \rho} - \bar{R} - (K^2-K_{\mu\nu}K^{\mu\nu}),
\end{equation}
in the Riemannian case. The first term vanishes by the Hamiltonian constraints and we end up with the same results as if we began with the Lagrangian.

It was shown in \cite{Henneaux:1979vn} that removing the terms that switches sign with signature results in a quantity that respects Carroll symmetries, i.e, $K^2-K_{\mu\nu}K^{\mu\nu}$. This is effectively the same as setting the signature of the manifold to zero. We will do the same procedure using the Lagrangian formalism in the following section in a different way, by considering the signature as a real parameter.

Contrary to the previously reviewed approaches, here we use the usual Levi-Civita connection. Expressing the Lagrangian in terms of the ADM variables, then setting the signature of the manifold to zero, we get a degenerate metric with the Carrollian theory (on a Carrollian manifold).

\subsection{Infinite scaling approach}
Starting from a Lorentzian theory and writing the metric in the same form as in the ZS approach but taking the limit of the norm of the normal vector to $-\infty$ results in a Galilean theory. Alternatively, starting with the Riemannian theory 
and taking the limit to $\infty$ results in the same Galilean theory; the difference would be only in an overall sign in the Lagrangian.

\subsection{Carroll algebra gauging}\label{CAG}
The gauging procedure for the Carroll algebra, the CAG approach, was done in \cite{Bergshoeff:2017btm,Hartong:2015xda,Figueroa-OFarrill:2022mcy}. We will briefly review it and show its equivalence with the PUL parametrization. The generators of the Carroll algebra are $H$ (time translation), $P_a$ (space translations), $G_a$ (Carroll boosts), and $J_{ab}$ (space rotations). We assign to each generator a gauge field as follows
\begin{equation}
\begin{aligned}
    H \to \tau_{\mu}, \hspace{10pt} & P_{a} \to e_{\mu}{}^a,  &  G_a \to \omega_{\mu}{}^a, & \hspace{10pt} J_{ab} \to \omega_{\mu}{}^{ab}.
    \end{aligned}
\end{equation}
The most general connection for the resulting geometry is
\begin{equation}
    \mathcal{A}_{\mu}= H\tau_{\mu} + P_ae_{\mu}{}^a + G_a\omega_{\mu}{}^a + \tfrac{1}{2} J_{ab}\omega_{\mu}{}^{ab}.
\end{equation}
Let us define an infinitesimal parameter for each gauge field by the virtue of an infinitesimal vector field $\xi^{\mu}$,
\begin{equation}
\begin{aligned}
    \zeta = \xi^{\mu}\tau_{\mu}, \hspace{10pt} & \zeta^{a}=\xi^{\mu}e_{\mu}{}^a, & \hspace{10pt} \lambda^{a}, \hspace{5pt} & \hspace{10pt} \lambda^{ab}.
    \end{aligned}
\end{equation}
Using these parameters we construct the infinitesimal transform parameter
\begin{equation}
    \Pi = H\zeta + P_a\zeta^a + G_a \lambda^a + \tfrac{1}{2}J_{ab}\lambda^{ab} = \xi^{\mu}\mathcal{A}_{\mu} + G_a \lambda^a + \tfrac{1}{2}J_{ab}\lambda^{ab}.
\end{equation}
The second equality follows from $\xi^{\mu}\omega_{\mu}^a=\xi^{\mu}\omega_{\mu\nu}^{ab}=0$. Let $\omega$ be a real parameter. Considering the geneators as rescaled Poincare generators, We notice that the rescaling of $\tau_{\mu}$ and $\zeta$ by $\omega^{-1}$ and of $\omega_{\mu}{}^a$ and $\lambda^a$ by $\omega$ leaves the connection $\mathcal{A}_{\mu} $ and the infinitesimal transformation parameter $\Pi$ invariant In the Carrollian limit (where $H$ and $G_a$ get rescaled by $\omega$). This consideration will be useful to define a metric usind the scaling parameter $\omega$. The field strength tensor (curvature) associated to $\mathcal{A}_{\mu}$ as 
\begin{equation}
    F_{\mu\nu} = \partial_{\mu}\mathcal{A}_{\nu} - \partial_{\nu}\mathcal{A}_{\mu} + [\mathcal{A}_{\mu}, \mathcal{A}_{\nu}] = H R_{\mu\nu}(H) + P_{a}R_{\mu\nu}{}^{a}(P)+ G_a R_{\mu\nu}{}^{a}(G) + \tfrac{1}{2}J_{ab}R_{\mu\nu}{}^{ab}(J),
\end{equation}
where $\partial_{\mu}$ is the partial derivative with respect to a generic coordinate system, and $ R_{\mu\nu}(H),R^a_{\mu\nu}(P),R^{ab}_{\mu\nu}(J), R^a_{\mu\nu}(G)$ are curvatures associated with the Carroll algebra generators which are given by
\begin{equation}\label{curvature}
    \begin{aligned}
        R_{\mu\nu}(H) &= 2 \partial_{[\mu}\tau_{\nu]} - 2 \omega_{[\mu}{}^ae_{\nu]a}, & \hspace{10pt}
        R_{\mu\nu}{}^{a}(P) &= 2 \partial_{[\mu}e_{\nu]}{}^a - 2 \omega_{[\mu}{}^{ab}\omega_{\nu]b},\\
        R_{\mu\nu}{}^{a}(G) &= 2 \partial_{[\mu}\omega_{\nu]}{}^a - 2\omega_{[\mu}{}^{ab}\omega_{\nu]b}, & \hspace{10pt}
        R_{\mu\nu}{}^{ab}(J) &= 2 \partial_{[\mu}\omega_{\nu]}{}^{ab} - 2 \omega_{[\mu}{}^a_c \omega_{\nu]}{}^{cb}.
    \end{aligned}
\end{equation}
These quantities can be used to construct the torsion and the Riemann tensor as shown in \cite{Lian:2021udx}.

\subsection{Galilei algebra gauging}
Let us also review the gauging procedure for the Galilei algebra, i.e., the GAG approach. The procedure is the same as the CAG procedure but with different generators \cite{Bergshoeff:2017btm}. We first assign a gauge field to each generator,
\begin{equation}
\begin{aligned}
  H \to \tau_{\mu}, & \hspace{10pt} P_a \to e_{\mu}{}^a, &  G_a \to \omega_{\mu}{}^a, & \hspace{10pt} J_{ab} \to \omega_{\mu}{}^{ab},  
  \end{aligned}
\end{equation}
where $G_a$ denote the Galilean boosts. The general connection and the infinitesimal transformation parameter are the same as in the Carrollian case. However, as in the Carrollian case, by considering the Galilean algebra generators as rescaled Poincare generators, the rescaling of $\tau_{\mu}$ and $\zeta$ by $\omega$, and of $\omega_{\mu}{}^a$ and $\lambda^a$ by ${\omega}^{-1}$ leaves the connection and the infinitesimal transformation parameter invariant in the Galilean limit. The curvatures are given by
\begin{equation}
\begin{aligned}
    R_{\mu\nu}(H) &= 2 \partial_{[\mu}\tau_{\nu]},&
    R_{\mu\nu}{}^a(P) &= 2\partial_{[\mu}e_{\nu]}{}^a - 2 \omega_{[\mu}{}^{ab}e_{\nu]b} - 2\omega_{[\mu}{}^a\tau_{\nu]},\\
    R_{\mu\nu}{}^a(G) &= 2\partial_{[\mu}\omega_{\nu]}{}^a - 2\omega_{[\mu}{}^{ab}\omega_{\nu]b},& \hspace{10pt}
    R_{\mu\nu}{}^{ab}(J) &= 2 \partial_{[\mu}\omega_{\nu]}{}^{ab} - 2 \omega_{[\mu}{}^{ac}\omega_{\nu]}{}^b_c,
    \end{aligned}
\end{equation}
where $\partial_{\mu}$ is the partial derivative with respect to a generic coordinate system.


\section{Comparison of non-Lorentzian methods}\label{sec3}
Let us demonstrate the equivalence between the various approaches mentioned above.

\subsection{Equivalence of PUL and ZS approaches}

Here, we rewrite the formulae used in the ADM decomposition with an extra real parameter $\epsilon$. This parameter should be understood as a modification to the normalization of the vector $n_{\mu}$ that is orthogonal to the spacelike slices, but not necessarily orthonormal, ${n_{\mu}n^{\mu}=\epsilon}$. The metric is Lorentzian for ${\epsilon < 0}$, Riemannian for ${\epsilon > 0}$, and degenerate for ${\epsilon = 0}$. Following the usual ADM decomposition procedure, we write the metric and its inverse in an adapted coordinate system satisfying ${\boldsymbol{\partial}_t=N\boldsymbol{n}+\boldsymbol{N}}$,
\begin{equation}
g_{\mu\nu} = \begin{bmatrix}
\epsilon(N^2+\epsilon N^iN_i) & -\epsilon N_i\\
-\epsilon N_i & h_{ij} 
\end{bmatrix},
\end{equation}
\begin{equation}
g^{\mu\nu} = \begin{bmatrix}
\tfrac{1}{\epsilon N^2} & \tfrac{N^j}{N^2} \\
\tfrac{N^i}{N^2} & h^{ij}+\epsilon\frac{N^iN^j}{N^2} 
\end{bmatrix},
\end{equation}
where $N$ is the lapse function, $N_i$ is the shift vector, and $h_{ij}$ is the induced metric. The vector $n_{\mu}$ in adapted coordinates reads
\begin{equation}\label{eq.1}
\begin{aligned}
    n_{\mu}=\left[-\epsilon N,0,0,0\right],& \hspace{10pt}
    n^{\mu}=\left[-\tfrac{1}{N},\epsilon \tfrac{N^i}{N}\right].
\end{aligned}
\end{equation}
The metric and its inverse can be written in terms of the orthogonal (co)vector, the induced metric, and its inverse as
\begin{equation}\label{eq.2}
 \begin{aligned}
    & g_{\mu\nu}= h_{\mu\nu} + \frac{1}{\epsilon}n_{\mu}n_{\nu}, & \hspace{10pt}
     g^{\mu\nu} =  h^{\mu\nu} + \frac{1}{\epsilon}n^{\mu}n^{\nu}.
 \end{aligned}
\end{equation}
One can show that $h_{\mu}^{\nu}h_{\nu}^{\rho}=h^{\rho}_{\mu}$, and $h^{\mu}_{\nu}=\delta_{\nu}^{\mu}-\tfrac{1}{\epsilon}n^{\mu}n_{\nu}$. The extrinsic curvature is then given by
\begin{equation}
    K_{\mu\nu}= h_{\mu}^{\alpha}h_{\nu}^{\beta}\nabla_{\alpha}n_{\beta} = - \nabla_{\nu}n_{\nu} + \tfrac{1}{\epsilon}n_{\mu}n^{\alpha}\nabla_{\alpha}n_{\nu}.
\end{equation}
It is easy to deduce that $\boldsymbol{K}=-\tfrac{1}{2}\pounds_{\boldsymbol{n}} \boldsymbol{h}$. We notice that although \eqref{eq.1} was written in some adapted coordinates the fact that $n_{\mu}$ has an overall factor of $\epsilon$ while $n^{\mu}=O(1)$ as $\epsilon \to 0$ does not change. This is because the dependency on the signature does not depend on the choice of the coordinate system. Thus, we can rescale $n_{\mu} = -\epsilon \tilde{n}_{\mu}$ where $\tilde{n}_{\mu} = O(1)$ when $\epsilon \to 0$. Writing \eqref{eq.2} in terms of the covector $\tilde{n}_{\mu}$ we get
\begin{equation}\label{eq.3}
 \begin{aligned}
    & g_{\mu\nu}= h_{\mu\nu} + \epsilon \tilde{n}_{\mu}\tilde{n}_{\nu}, & \hspace{10pt}
     g^{\mu\nu} =  h^{\mu\nu} + \frac{1}{\epsilon}n^{\mu}n^{\nu},
 \end{aligned}
\end{equation}
with
\begin{equation}
    \begin{aligned}
\tilde{n}_{\mu}h^{\mu\nu}=n^{\mu}h_{\mu\nu}=0, & \hspace{10pt} \hspace{5pt}\tilde{n}_{\mu}n^{\mu}=-1, & \hspace{10pt} h_{\mu\nu}h^{\nu\rho} - \tilde{n}_{\mu}n^{\rho} = \delta^{\rho}_{\mu}.
    \end{aligned}
\end{equation}
 and 
\begin{equation}
    K_{\mu\nu}= \epsilon  \big(\nabla_{\mu}\tilde{n}_{\nu}+\tilde{n}_{\mu}n^{\alpha}\nabla_{\alpha}\tilde{n}_{\nu}\big).
\end{equation}
This is the PUL form that appeared in \cite{Hansen:2021fxi} with $\epsilon = -c^2$. Thus, the limit $\epsilon \to 0$ gives the same result as the limit $c \to 0$.\footnote{A Galilean analogue of this limit is to take $\epsilon \to -\infty$. This limit gives  correct Galilean expansions. Given that the parameter $\epsilon$ is related to the norm of the normal vector. The Galilean limit corresponds to infinite normalization of this vector.} This means that the two methods are equivalent. It was shown that the result would be a Carrollian manifold in \cite{Henneaux:1979vn}.

We now present the explicit calculations for GR. The resulting formulae are the same as the formulae derived using the PUL approach up to renaming of quantities. We can write the extrinsic curvature in terms of partial derivatives instead of covariant derivatives as follows 
\begin{equation}\label{extrinsic curvature}
    K_{\mu\nu} = -\tilde{K}_{\mu\nu} + \epsilon \big(\partial_{[\mu}\tilde{n}_{\nu]} + n^{\sigma} n_{[\nu}\partial_{\sigma}n_{\nu]}\big),
\end{equation}
where $\tilde{K}_{\mu\nu} = 2n^{\sigma}\partial_{(\mu}h_{\nu)\sigma} - n^{\sigma}\partial_{\sigma}h_{\mu\nu}$, and $\partial_{\mu}$ is the partial derivative with respect to a generic coordinate system.
Assuming that the quantities $h_{\mu\nu}, h^{\mu\nu}, \tilde{n}_{\mu}, n^{\mu}$ are analytic, if we would to expand them in powers of $\epsilon$ similar to the expansions done in the PUL approach. The LO defines a Carrollian theory (respects Carroll symmetries), while the NLO breaks the symmetry unless we set all the higher order fields to zero i.e, to assume that all the quantities do not depend on $\epsilon$. This is similar to the truncation procedure in \cite{Hansen:2021fxi} to get the magnetic limit from the NLO expansion. It is worth mentioning that we use the usual Levi-Civita connection not the connection in \cite{Hansen:2021fxi}. Although the covariant derivative does not preserve $n^{\mu}$ and $h_{\mu\nu}$, we still get Carrollian theories at the end.

Having shown that, we write Gauss relation and its contractions in terms of the parameter $\epsilon$. Let us begin with the Gauss relation,
\begin{equation}\label{gauss eq}
    h_{\alpha}^{\mu}h_{\beta}^{\nu}h_{\rho}^{\gamma}h_{\sigma}^{\delta}R^{\rho}{}_{\sigma\mu\nu}= \bar{R}^{\gamma}{}_{\delta \alpha \beta}-\tfrac{1}{\epsilon}\big(K^{\gamma}_{\alpha}K_{\delta \beta}-K^{\gamma}_{\beta}K_{\delta \alpha}\big),
\end{equation}
where $R^{\rho}{}_{\sigma\mu\nu}$ is the 4-dimensional Riemann tensor and  $\bar{R}^{\gamma}{}_{\delta \alpha \beta}$ is the 3-dimensional Riemann tensor on a spacelike slice. Contracting $\gamma$ and $\alpha$ and using $ h_{\alpha}^{\mu}h_{\rho}^{\alpha}=h_{\rho}^{\mu}=\delta_{\rho}^{\mu}-\tfrac{1}{\epsilon}n^{\mu}n_{\rho}$, we get the contracted Gauss relation
\begin{equation}    
    h_{\beta}^{\nu}h_{\alpha}^{\sigma}R_{\sigma\nu}-\tfrac{1}{\epsilon}\big(h_{\beta \mu}h_{\alpha}^{\nu}n^{\rho}n^{\sigma}{}_{(4)}{R^{\mu}{}_{\rho\nu\sigma}}\big) = \bar{R}_{\alpha\beta} - \tfrac{1}{\epsilon}\big(KK_{\alpha\beta}-K_{\alpha}^{\delta}K_{\delta \beta}\big), 
\end{equation}
\begin{equation}
    R-\tfrac{1}{\epsilon}\big(2n^{\mu}n^{\nu}R_{\mu\nu}\big) = \bar{R}-\tfrac{1}{\epsilon}\big(K^2-K^{\mu\nu}K_{\mu\nu}\big),
\end{equation}
where $R$ is the 4-dimensional Ricci scalar and $\bar{R}$ is the 3-dimensional Ricci scalar for a spacelike slice. After some algebra we get
\begin{equation}
    R = \bar{R} - \tfrac{1}{\epsilon}\big(K^2-K^{\mu\nu}K_{\mu\nu} - 2\nabla_{\mu}A^{\mu}\big),
\end{equation}
where $A^{\mu} = -n^{\mu}\nabla_{\nu}n^{\nu} + n^{\nu}\nabla_{\nu}n^{\mu}$. Using \eqref{extrinsic curvature} we see that ${K^2=\tilde{K}^2}$, ${K_{\mu\nu}K^{\mu\nu}=\tilde{K}_{\mu\nu}\tilde{K}^{\mu\nu} + \epsilon^2(\boldsymbol{\mathrm{d}}\boldsymbol{n})^2}$ where $\boldsymbol{\mathrm{d}}\boldsymbol{n}$ is the exterior derivative of $n_{\mu}$. Putting all together, we get
\begin{equation}\label{expansion of Ricci scalar}
     R = -\tfrac{1}{\epsilon}\big(\tilde{K}^2 - \tilde{K}_{\mu\nu}\tilde{K}^{\mu\nu} -2 \nabla_{\mu}A_1^{\mu}\big) + \bar{R}-2\nabla_{\mu}A_2^{\mu} + \epsilon (\boldsymbol{\mathrm{d}}\boldsymbol{n})^2,
\end{equation}
where $A_1^{\mu}= - n^{\mu}\partial_{\nu}n^{\nu} + n^{\nu}\partial_{\nu}n^{\mu}-\tfrac{1}{2}n^{\mu}n^{\sigma}h^{\alpha\rho}\big(2\partial_{(\alpha}h_{\sigma)\rho} - \partial_{\rho}h_{\alpha \sigma}\big) + h^{\mu\rho}n^{\sigma}n^{\nu}\partial_{\nu}h_{\sigma \rho} - \tfrac{1}{2}n^{\sigma}n^{\nu}h^{\mu\rho}\partial_{\rho}h_{\sigma \nu}$,
and $A_2^{\mu} = -h^{\mu\rho}n^{\sigma}\partial_{\sigma}n_{\rho}$, where $\partial_{\mu}$ is the partial derivative with respect to a generic coordinate system. This formula is the same as the PUL parameterization of the Ricci scalar up to change of connections and renaming quantities.

\subsection{Equivalence of PNR and IS approaches}
Similar to the previous subsection, starting from \eqref{eq.3} and taking the limit $\epsilon \to -\infty$, results in the metric in \cite{Hansen:2020wqw} in the PNR parameterization. Thus, all the curvature tensors in both approaches must be equivalent. All the formulae afterwards are valid for the Galilean limit provided (in terms of a different connection) we take the limit $\epsilon \to -\infty$. The limit $\epsilon \to \infty$ gives the same formulae except for an overall negative sign.

\subsection{Equivalence of PUL and CAG approaches}
Let us now reconstruct the PUL parametrization from the CAG procedure. 
Using the procedure delineated in Sec.~\ref{CAG}, we define inverses to the gauge fields vielbein ($\tau_{\mu}$ and $e_{\mu}{}^a$) by
\begin{equation}
    \begin{aligned}
        v^{\mu}\tau_{\mu}=-1, & \hspace{10pt} v^{\mu}e_{\mu}{}^a= \tau_{\mu}e^{\mu}{}_b=0, \hspace{10pt} & e_{\mu}{}^a e^{\mu}{}_b=\delta_b^a, \hspace{10pt} & e_{\mu}{}^a e^{\nu}{}_a= \delta_{\mu}^{\nu}-\tau_{\mu}v^{\nu}.
    \end{aligned}
\end{equation}
Performing the redefinitions in \cite{Bergshoeff:2017btm}, where $\tau_{\mu}$ , $\omega_{\mu}{}^a$, $\zeta$, and $\lambda^a$ are rescaled by $\omega^{-1}$ and taking the limit $\omega \to \infty$, leaving $\mathcal{A}_{\mu}$ and $\Pi$ invariant, we can write the metric and its inverse as

\begin{equation}
\begin{aligned}
    g_{\mu\nu} = \tfrac{-1}{\omega}\tau_{\mu}\tau_{\nu}+ e_{\mu}{}^a e_{\nu}{}^b \delta_{ab}, \hspace{10pt} & g^{\mu\nu} = -\omega v^{\mu}v^{\nu}+ e^{\mu}{}_a e^{\nu}{}_b \delta^{ab}.
    \end{aligned}
\end{equation}
This form is the same form in the PUL approach with ${\omega = c^{-2}}$. It follows from this fact is that all curvature tensors will be the same in the two approaches. Thus, any gravity theory will be expanded similarly, i.e, the two approaches are equivalent.

We now derive the Carroll compatible connection presented in \cite{Hansen:2021fxi} from the CAG approach. To do that, let us define a covariant derivative that are compatible with the veilbein \cite{Bergshoeff:2017btm} by
\begin{equation}
    \begin{aligned}
      D_{\mu}\tau_{\nu} &= \partial_{\mu}\tau_{\nu} + C^{\rho}_{\mu\nu}\tau_{\rho} - \omega_{\mu a}e_{\nu}{}^a = 0,& \hspace{10pt}
      D_{\mu}e_{\nu}{}^a &= \partial_{\mu}e_{\nu}{}^a - C^{\rho}_{\mu\nu}e_{\rho}{}^a - \omega_{\mu}{}^a_b e_{\nu}^b = 0,\\
      D_{\mu}v^{\nu} &= \partial_{\mu}v^{\nu} + C_{\mu \rho}^{\nu}v^{\rho}= 0,& \hspace{10pt}
      D_{\mu}e^{\nu}{}_a &= \partial_{\mu}e^{\nu}{}_a + C^{\nu}_{\mu\rho}e^{\rho}{}_a - \omega_{\mu a} v^{\nu} - \omega_{\mu}{}_a^b e^{\nu}{}_b = 0, 
    \end{aligned}
\end{equation}
where $C^{\rho}_{\mu\nu}$ is the Carroll compatible connection, $\omega_{\mu a}$ and $\omega_{\mu}{}^a_b$ act as spin connections, and $\partial_{\mu}$ is the partial derivative with respect to a generic coordinate system. Defining $h_{\mu\nu} = e_{\mu}{}^ae_{\nu}{}^b\delta_{ab}$ and solving for $C^{\rho}_{\mu\nu}$, we see that the simplest connection to satisfy the conditions is
\begin{equation}
    C^{\rho}_{\mu\nu} = -v^{\rho} \partial_{(\mu}\tau_{\nu)} - v^{\rho} \tau_{(\mu}\pounds_{\boldsymbol{v}}\tau_{\nu)} + \tfrac{1}{2}h^{\rho \sigma}\big(\partial_{\mu}h_{\nu \sigma} + \partial_{\nu}h_{\mu \sigma} - \partial_{\sigma}h_{\mu\nu}\big) + \tfrac{1}{2} h^{\rho \sigma}\tau_{\nu}\pounds_{\boldsymbol{v}} h_{\mu\sigma}.
\end{equation}
(For more general connections see \cite{Hartong:2015xda}.) This procedure is the generalized version of the one in appendix B of \cite{Hansen:2021fxi}. It is easy to see from this point using \eqref{curvature} that the torsion and the curvature tensors match that of the PUL ones. For example, the Riemann tensor is given by
 \begin{equation}
     R^{\rho}{}_{\mu\nu\sigma} = v^{\rho}e_{\sigma a}R_{\mu\nu}{}^a(G) + e^{\rho}{}_ae_{\sigma b}R_{\mu\nu}{}^{ab}(J),
 \end{equation}
which, upon calculating explicitly, gives the same form as (2.18) in \cite{Tadros:2023teq}.

\subsection{Equivalence of PNR and GAG approaches}
In a similar fashion to the previous section, we show that the GAG procedure leads to the same theory as the PNR parametrization. This is inspired by the Carroll-Galilei duality which was recently found to hold to the level of gauge fields not only the algebra \cite{Bergshoeff:2020xhv,Bergshoeff:2023rkk}.
Similar to the previous section, the metric can be written as
\begin{equation}
    \begin{aligned}
       g_{\mu\nu} = -\omega \tau_{\mu}\tau_{\nu}+ e_{\mu}{}^a e_{\nu}{}^b \delta_{ab}, \hspace{10pt} & g^{\mu\nu} = -\tfrac{1}{\omega} v^{\mu}v^{\nu}+ e^{\mu}{}_a e^{\nu}{}_b \delta^{ab}.
    \end{aligned}
\end{equation}
This is the same form as in the PNR parameterization with $\omega = c^2$. We can also define a Galilei compatible covariant derivative as 
\begin{equation}
    \begin{aligned}
        D_{\mu}\tau_{\nu} &= \partial_{\mu}\tau_{\nu} - C^{\rho}_{\mu\nu}\tau_{\rho} = 0,&
        D_{\mu}e_{\nu}{}^a &= \partial_{\mu}e_{\nu}{}^a - C^{\rho}_{\mu\nu}e_{\rho}{}^a - \omega_{\mu}{}^a\tau_{\nu} - \omega_{\mu}{}^a_b e_{\nu}{}^b = 0,\\
        D_{\mu}v^{\nu} &= \partial_{\mu}v^{\nu} + C^{\nu}_{\mu\rho}v^{\rho} - \omega_{\mu}{}^a e^{\nu}{}_a = 0,& \hspace{10pt}
        D_{\mu}e^{\nu}{}_a &= \partial_{\mu}e^{\nu}{}_a + C_{\mu\rho}^{\nu}e^{\rho}{}_a - \omega_{\mu}{}^b{}_a e^{\nu}{}_b = 0,
    \end{aligned}
\end{equation}
where $C^{\rho}_{\mu\nu}$ is a Galilei compatible connection, $\omega_{\mu}{}^a$ and $\omega_{\mu}^{ab}$ act as spin connections, and $\partial_{\mu}$ is the partial derivative with respect to a generic coordinate system. Defining ${h_{\mu\nu} = e_{\mu}{}^ae_{\nu}{}^b\delta_{ab}}$, we can see that the simplest connection that satisfies the above conditions is
\begin{equation}
    C^{\rho}_{\mu\nu} = -v^{\rho}\partial_{\mu}\tau_{\nu} + \tfrac{1}{2}h^{\rho \sigma}\big(2\partial_{(\mu}h_{\nu)\sigma} - \partial_{\sigma}h_{\mu\nu}\big),
\end{equation}
 which is the same connection suggested in \cite{Hansen:2020wqw}. The respective Riemann tensor is given by
 \begin{equation}
     R^{\rho}{}_{\mu\nu\sigma} = e_{\sigma a}e^{\rho}{}_{b}R_{\mu\nu}{}^{ab}(J) - e^{\rho}{}_{a}\tau_{b}R_{\mu\nu}{}^a(G).
 \end{equation}


\section{Applications to HDG} \label{sec6}
Having established that all methods of taking the Galilean and Carrollian limits are equivalent, we derive an algorithm to calculate the $n-$th order of the Galilean and Carrollian expansions of a generic HDG theory. 

\subsection{$f(R)$ gravity}
Here we derive the Galilean and the Carrollian limits for a general $f(R)$ gravity theory to the LO and NLO. We introduce an algorithm for finding any order of the expansions using combinatorial arguments. The Lagrangian for $f(R)$ gravity is 
\begin{equation}\label{f(R) Lagrangian}
    \mathcal{L} = f(R),
\end{equation}
where $R$ is the Ricci scalar and $f$ is an analytic function. We expand $f$ as a power series in integer powers of $R$, with coefficients $g_i$:
\begin{equation}
    f(R)= \sum_{i=0}^{\infty}g_i R^i,
\end{equation}
where $g_i$ are constant coefficients that are independent of $c$.

Since all methods of constructing the Galilean and Carrollian expansions are equivalent as shown in previous sections, we can write the Ricci scalar as a generic expansion as follows
\begin{equation}\label{Ricci scalar expansion}
    R = \tfrac{1}{c^2}R_1 + R_2 + c^2 R_3,
\end{equation}
where $R_1,R_2,R_3$ are the expansion terms shown explicitly in (3.2) in \cite{Hansen:2021fxi}; they are also displayed in \eqref{ricci scalar}. (The use of the connection does not affect the from of the expansion as shown in Sec.~\ref{sec3}, different connections will result in different forms for $R_1$,$R_2$,$R_3$).

The constants $g_i$ are considered as functions of $c$ (not necessarily regular). Writing
\begin{equation}\label{expansion of g}
    g_i(c)=\sum_{j=-\infty}^{\infty}a_{ij}c^j,
\end{equation} 
where $a_{ij}$ are constants without any $c$ dependency, and expanding $f(R)$, we get
\begin{equation}
    f(R) = \sum_{i=0}^{\infty}\sum_{j=-\infty}^{\infty}a_{ij}\sum_{\mathclap{\substack{n,m,k \geq 0 \\ n+m+k=i}}}c^{2k-2n+j}\frac{i!}{n!m!k!} R_1^n R_2^m R_3^k.
\end{equation}
Here, we need to be extra careful since $j$ can be unbounded regardless of the value of $i$, and appears in the power of $c$. For a theory to have consistent Galilean and Carrollian limits, the quantity ${2k-2n+j}$ must be bounded from above and below.

\subsubsection{Carrollian limit}

Let us denote $l=\min(2k-2n+j)$. The electric limit's Lagrangian is given by 
\begin{equation}\label{electric lagrangian}
\begin{aligned}
\mathcal{L}_{\textrm{el}} &=c^l\sum_{i=0}^{\infty}\sum_{j=-\infty}^{\infty}a_{ij}\sum_{\mathclap{\substack{n,m,k \geq 0 \\ n+m+k=i \\ 2k-2n+j=l}}} \hspace{5pt} \frac{i!}{n!m!k!}R_1^n R_2^m R_3^k 
\\
&= c^l \sum_{i=0}^{\infty}\sum_{j=-\infty}^{\infty}a_{ij}\sum_{\mathclap{\substack{n,m \geq 0\\ 2n+m+\frac{l-j}{2}=i}}} \hspace{5pt} \frac{i!}{n!m!\big(n+\frac{l-j}{2}\big)!}R_1^n R_2^m R_3^{n+\tfrac{l-j}{2}},
\end{aligned}
\end{equation}
while the magnetic limit's Lagrangian reads
\begin{equation}\label{magnetic lagrangian}
\begin{aligned}
    \mathcal{L}_{\textrm{mag}} &= c^{l+2} \sum_{i=0}^{\infty}\sum_{j=-\infty}^{\infty}a_{ij}\sum_{\mathclap{\substack{n,m,k \geq 0 \\ n+m+k=i \\ 2k-2n+j=l+2}}} \hspace{5pt} \frac{i!}{n!m!k!}R_1^n R_2^m R_3^k
    \\
    &= c^{l+2} \sum_{i=0}^{\infty}\sum_{j=-\infty}^{\infty}a_{ij}\sum_{\mathclap{\substack{n,m \geq 0\\ 2n+m+\frac{l+2-j}{2}=i}}} \hspace{5pt} \frac{i!}{n!m!\big(n+\tfrac{l+2-j}{2}\big)!}R_1^n R_2^m R_3^{n+\tfrac{l+2-j}{2}}.
    \end{aligned}
\end{equation}
As a specific example we now compute the Carrollian limits of the theories $f(R) = R + c^N \alpha R^2$ where $\alpha$ is a constant with no $c$ dependency. The Carrollian limit of these type of theories was computed in \cite{Tadros:2023teq}. The electric limit can be computed from \eqref{electric lagrangian} by noticing that all the coefficients $a_{ij}$ are zero except $a_{10}=1$ and $a_{2N}=\alpha$. In this case $i=1,2$ and $j=0,N$.

The term where $i=1$ and $j=0$ have $l=-2$, $m=0$ and $n=1$, while the term where $i=2$ and $j=N$ has $l=N-4$, $m=0$ and $n=2$. Putting all together we get
\begin{equation}
    \mathcal{L}_{\textrm{el}}= c^{-2}R_1 + 2\alpha c^{N-4}R_1^2.
\end{equation}
In order for the second term to couple to some higher order of the first term, we must set $N \geq 2$ otherwise, it will be pure quadratic gravity, i.e., not a modification of GR. This is the same condition derived in \cite{Tadros:2023teq}.
The magnetic limit can be derived from \eqref{magnetic lagrangian} by noting that the first term has $i=1$, $j=0$, $l=-2$. Solving $2n+m+\frac{l+2-j}{2}=i$, we get $m=1$ and $n=0$. The second term has $i=2$, $j=N$ and $l=N-4$. Again, by solving $2n+m+\tfrac{l+2-j}{2}=i$, we get $m=n=1$. Putting all together we obtain
\begin{equation}
    \mathcal{L}_{\textrm{mag}} = R_2 + 2\alpha c^{N-2}R_1R_2,
\end{equation}
with the condition $N \geq 2$ to get a GR modification.

For the case where $f(R)$ does not depend on $c$ explicitly, the electric limit's Lagrangian is given by
\begin{equation}\label{electric limit}
    \begin{aligned}
        \mathcal{L}_{\textrm{el}} &= c^{l}\sum_{i=0}^{\infty}g_i \sum_{\mathclap{\substack{n,m,k \geq 0 \\ n+m+k=i \\ 2k-2n=l}}} \hspace{5pt}  \frac{i!}{n!m!k!} R_1^n R_2^m R_3^k 
        \\
        &= c^{l} \sum_{i=0}^{\infty}g_i \sum_{\mathclap{\substack{n,m \geq 0\\ 2n+m+\frac{l}{2}=i}}} \hspace{5pt}  \frac{i!}{n!m!\big(\frac{l}{2}+n\big)!} R_1^n R_2^m R_3^{\frac{l}{2} + n},
    \end{aligned}
\end{equation}
and the magnetic limit's Lagrangian is
\begin{equation}\label{magnetic limit}
    \begin{aligned}
        \mathcal{L}_{\textrm{mag}} &= c^{l+2} \sum_{i=0}^{\infty}g_i \sum_{\mathclap{\substack{n,m,k \geq 0\\ n+m+k=i \\ 2k-2n=l+2}}} \hspace{5pt} \frac{i!}{n!m!k!} R_1^n R_2^m R_3^k 
        \\
        &= c^{l+2} \sum_{i=0}^{\infty}g_i \sum_{\mathclap{\substack{n,m \geq 0 \\ 2n+m+\frac{l}{2}=i-1}}} \hspace{5pt} \frac{i!}{n!m!\big(\frac{l}{2} + n + 1\big)!} R_1^n R_2^m R_3^{\tfrac{l}{2} + n + 1}.
    \end{aligned}
\end{equation}
As consistency check, let us calculate the electric and magnetic limits of GR, $f(R)=R$. In this case $g_i = 0$ except for $g_1 = 1$; thus
\begin{equation}
    \mathcal{L}_{\textrm{el}} = c^{l}\sum_{\mathclap{\substack{n,m \geq 0\\ 2n+m+\frac{l}{2}=1}}} \hspace{5pt} \frac{1}{n!m!\big(\frac{l}{2} + n\big)!} R_1^n R_2^m R_3^{\tfrac{l}{2} + n}.
\end{equation}
Here the minimum value of $2k-2n$ is $-2$ i.e,  $l = -2$ (by choosing $k=0, n=1$), and since $2n+m+\frac{l}{2}=1$ therefore $m=0$, substituting in the formula we get
\begin{equation}
    \mathcal{L}_{\textrm{el}} = c^{-2}R_1.
\end{equation}
The magnetic limit can be derived similarly from
\begin{equation}
    \mathcal{L}_{\textrm{mag}} = c^{l} \sum_{\mathclap{\substack{n,m \geq 0\\ 2n+m+\frac{l}{2}=0}}} \hspace{5pt} \frac{1}{n!m!\big(\frac{l}{2} + n + 1\big)!} R_1^n R_2^m R_3^{\tfrac{l}{2} + n + 1},
\end{equation}
by solving $2n+m = 1$, with in this case $n$ must be zero and $m=1$. Thus,
\begin{equation}
\mathcal{L}_{\textrm{mag}} = R_2.
\end{equation}
The $n-$th order in the Carrollian expansion can be derived from the same minimization problem by replacing $l$ with $l+2n$.

\subsubsection{Galilean limit}

The Galilean limit uses the same formulae except for setting $l=\max(2k-2n+j)$. The LO takes the form
\begin{equation}
\begin{aligned}
\mathcal{L}_{\textrm{NC}} &=c^l\sum_{i=0}^{\infty}\sum_{j=-\infty}^{\infty}a_{ij}\sum_{\mathclap{\substack{n,m,k \geq 0 \\ n+m+k=i \\ 2k-2n+j=l}}} \hspace{5pt} \frac{i!}{n!m!k!}R_1^n R_2^m R_3^k 
\\
&= c^l \sum_{i=0}^{\infty}\sum_{j=-\infty}^{\infty}a_{ij}\sum_{\mathclap{\substack{n,m \geq 0\\ 2n+m+\frac{l-j}{2}=i}}} \hspace{5pt} \frac{i!}{n!m!\big(n+\frac{l-j}{2}\big)!}R_1^n R_2^m R_3^{n+\frac{l-j}{2}},
\end{aligned}
\end{equation}
which is similar to the Carrollian case but with $l$ being the maximum rather than the minimum. The NLO reads
\begin{equation}
    \begin{aligned}
    \mathcal{L}_{\textrm{TNC}} = c^{l-2} \sum_{i=0}^{\infty}\sum_{j=-\infty}^{\infty}a_{ij}\sum_{\mathclap{\substack{n,m \geq 0\\ 2n+m+ \frac{l-j}{2}=i+1}}} \hspace{5pt} \frac{i!}{n!m!\big(n-1+\frac{l-j}{2}\big)!}R_1^n R_2^m R_3^{n-1 +\tfrac{l-j}{2}}.
    \end{aligned}
\end{equation}
We will derive the Galilean limit of $f(R)=R+c^N \alpha R^2$ as an example. As mentioned in the previous section, all coefficients $a_{ij}$, except for $a_{10}=1$ and $a_{2N}=\alpha$, vanish. The first term in the Galilean limit has $i=1$, $j=0$ and $l=2$ i.e, $n=0$. Solving $2n+m+\tfrac{l-j}{2}=i$ leads to $m=0$. The second term has $i=2$, $j=N$, $l=N+4$ i.e, $n=0$. Solving $2n+m+\tfrac{l-j}{2}=i$ gives $m=0$. Putting all together, we get
\begin{equation}
    \mathcal{L}_{\textrm{NC}} = c^2R_3 + 2\alpha c^{N+4}R_3^2.
\end{equation}
For this theory to be a modification of the Galilean limit of GR at some order, we must impose the condition $N \leq -2$. Otherwise it would be a theory of pure quadratic gravity. 

The first terms in the NLO has $i=1$, $j=0$, $l=2$. Solving $2n+m+\tfrac{l-j}{2}=i+1$, we get $n=0$, $m=1$. The second term has $i=2$, $j=N$ and $l= N+4$. Solving $2n+m+\tfrac{l-j}{2}=i+1$, we get $n=0$, $m=1$. Putting all together we get
\begin{equation}
    \mathcal{L}_{\textrm{TNC}} = R_2 + 2\alpha c^{N+2} R_2R_3,
\end{equation}
with the condition $N \leq -2$ to get a GR modification. From this condition and the condition for the Carrollian case, we see that no $f(R)$ theory with a polynomial function $f$ can be a modification of GR in both the Galilean and the Carrollian regimes at the same time.

For the case where $f(R)$ does not depend on $c$ explicitly, the LO takes the form \begin{equation}
    \begin{aligned}
        \mathcal{L}_{\textrm{NC}} &= c^{l}\sum_{i=0}^{\infty}g_i \sum_{\mathclap{\substack{n,m,k \geq 0 \\ n+m+k=i \\ 2k-2n=l}}} \hspace{5pt}  \frac{i!}{n!m!k!} R_1^n R_2^m R_3^k 
        \\
        &= c^{l} \sum_{i=0}^{\infty}g_i \sum_{\mathclap{\substack{n,m \geq 0\\ 2n+m+\frac{l}{2}=i}}} \hspace{5pt}  \frac{i!}{n!m!\big(\frac{l}{2}+n\big)!} R_1^n R_2^m R_3^{\tfrac{l}{2} + n}.
    \end{aligned}
\end{equation}
The NLO, however, is a little different, but the general formula can be derived in a similar way to the Carrollian case,
\begin{equation}
    \mathcal{L}_{\textrm{TNC}}= c^{l-2} \sum_{i=0}^{\infty}g_i \sum_{\mathclap{\substack{n,m \geq 0 \\ 2n+m+\frac{l}{2}=i+1}}} \hspace{5pt} \frac{i!}{n!m!\big(\frac{l}{2} + n - 1\big)!} R_1^n R_2^m R_3^{\frac{l}{2} + n - 1}.
\end{equation}
Let us again check these expressions by applying to GR again. The LO reduces to
\begin{equation}
    \mathcal{L}_{\textrm{NC}} = c^{l} \sum_{\mathclap{\substack{n,m \geq 0\\ 2n+m+\frac{l}{2}=1}}} \hspace{5pt} \frac{1}{n!m!\big(\frac{l}{2} + n\big)!} R_1^n R_2^m R_3^{\tfrac{l}{2} + n},
\end{equation}
where $l=2$. Then by $2n+m+(l/2)=1$, we get $n=m=0$, so the LO is
\begin{equation}
    \mathcal{L}_{\textrm{NC}}= c^2R_3.
\end{equation}
The NLO can be derived from
\begin{equation}
    \mathcal{L}_{\textrm{TNC}} = c^{l-2} \sum_{\mathclap{\substack{n,m \geq 0\\ 2n+m+\frac{l}{2}=i+1}}}\hspace{5pt}  \frac{i!}{n!m!\big(\frac{l}{2} + n - 1\big)!} R_1^n R_2^m R_3^{\tfrac{l}{2} + n - 1}.
\end{equation}
Here we have to solve $2n+m+1=2$. The solution is $n=0$ and $m=1$, and we get
\begin{equation}
    \mathcal{L}_{\textrm{TNC}} = R_2.
\end{equation}
The $n-$th order in the Galilean expansion can be derived from the same maximization problem by replacing $l$ with $l-2n$.

\subsection{$f(g_{\mu\nu},R_{\mu\nu\sigma\rho})$ gravity}
We first consider gravity theories of the form $f(g_{\mu\nu},R_{\mu\nu\sigma\rho})$ where $f$ is polynomial. We consider a generic term in the Lagrangian to apply the algorithm, the $n$-th term in the Carrollian or Galilean expansions is the sum of the contributions of all terms. A generic term in the Lagrangian can be written as
\begin{equation}\label{HDG Lagrangian}
    \mathcal{L} = \alpha \mathcal{R}^a \Bar{\mathcal{R}}^b,
\end{equation}
where $\mathcal{R}$ can be the Riemann tensor, the Ricci tensor with all indices down, or the Ricci scalar, $\Bar{\mathcal{R}}$ can be the same tensors with all indices up, and $a$, $b$ are positive numbers such that the Lagrangian terms are scalars. Using eq.(2.8) in \cite{Tadros:2023teq} and raising/lowering the indices, we can write the generic expansions
\begin{equation}\label{HDG expansion}
    \begin{aligned}
  \mathcal{R} &= \tfrac{1}{c^2} \mathcal{R}_1 + \mathcal{R}_2 + c^2 \mathcal{R}_3 + c^4\mathcal{R}_4,& \hspace{10pt}
  \Bar{\mathcal{R}} &= \tfrac{1}{c^4}\Bar{\mathcal{R}}_1 + \tfrac{1}{c^2}\Bar{\mathcal{R}}_2 +\Bar{\mathcal{R}}_3 + c^2\Bar{\mathcal{R}}_4.
    \end{aligned}
\end{equation}
The explicit expressions are in the appendix. Raising the expansions to the powers of $a$, $b$, and $c$, respectively, we obtain
\begin{equation}
    \begin{aligned}
        \mathcal{R}^a &= \sum_{\mathclap{\substack{\sum\limits_{i=0}^4k_i=a \\ k_i \geq 0}}} \frac{a!}{k_1!...k_4!} \prod_{t=1}^4 c^{2k_t(t-2)} \big(\mathcal{R}_t\big)^{k_t},& \hspace{10pt}
        \bar{\mathcal{R}}^b &= \sum_{\mathclap{\substack{\sum\limits_{j=0}^4k_j=b \\ k_j \geq 0}}} \frac{b!}{\Bar{k}_1!...\Bar{k}_4!} \prod_{\Bar{t}=1}^4 c^{2\Bar{k}_{\Bar{t}}(\bar{t}-3)} \big(\Bar{\mathcal{R}}_{\Bar{t}}\big)^{\Bar{k}_{\Bar{t}}}.
    \end{aligned}
\end{equation}
Assuming $\alpha = c^N \alpha'$ where $\alpha'$ has no $c$ dependency and substituting in the Lagrangian, we find
\begin{equation}
    \begin{aligned}
       \mathcal{L} = \alpha' \sum_{\mathclap{\substack{\sum\limits_{i=0}^4k_i=a \\ k_i \geq 0}}} \frac{a!}{k_1!...k_4!}\hspace{5pt}\sum_{\mathclap{\substack{\sum\limits_{j=0}^4\Bar{k}_j=b \\ k_j \geq 0}}} \frac{b!}{\Bar{k}_1!...\Bar{k}_4!}
    \prod_{t=1}^4\prod_{\Bar{t}=1}^4 c^{2k_t(t-2)+2\Bar{k}_{\Bar{t}}(\bar{t}-3) +N}(\mathcal{R})^{k_t}(\Bar{\mathcal{R}})^{\Bar{k}_{\Bar{t}}}.
    \end{aligned}
\end{equation}
The $n$-th order in the Galilean or the Carrollian limits can be obtained by solving a numerical optimization problem as we will demonstrate below.

\subsubsection{Galilean limit}
The LO of the Galilean limit is given by the solution of the optimization problem
\begin{equation}
        z=\max\bigg[\sum_{t=1}^4 2k_t(t-2)+2\sum_{\Bar{t}=1}^4 \Bar{k}_{\Bar{t}}(\bar{t}-3) +N\bigg],
\end{equation}
subjected to the constraints
\begin{equation}
    \begin{aligned}
        \sum_{i=0}^4k_i=a, & \hspace{10pt} \sum_{j=0}^4\Bar{k}_j=b, & k_i,\Bar{k}_j \geq 0.
    \end{aligned}
\end{equation}
  This is a constrained numerical optimization problem that can be solved by the simplex method \cite{borgwardt2012simplex} for $k_i$, $\Bar{k}_j$ given that $a$, $b$ are known. The solution is $z=4a + 4b + 2 +N$. Thus, for this term to be a modification of GR, the condition on $N$ is $N \leq -4a - 4b$. If such term exists in the Lagrangian then the Lagrangian is a modification of the Galilean limit of GR. The $n$-th order can be deduced by modifying the constrained optimization problem by $z \to z-2n$.

  \subsubsection{Carrollian limit}
  The LO is given by the solution of the constrained optimization problem
\begin{equation}
        z=\min\bigg[2\sum_{t=1}^4k_t(t-2)+2\sum_{\Bar{t}=1}^4\Bar{k}_{\Bar{t}}(\bar{t}-3) + N\bigg],
\end{equation}
subjected to the constraints
\begin{equation}
    \begin{aligned}
        \sum_{i=0}^4k_i=a, & \hspace{10pt} \sum_{j=0}^4\Bar{k}_j=b, & k_i,\Bar{k}_j \geq 0,
    \end{aligned}
\end{equation}
which is again solvable by the simplex method given that $a,b$ are known. The solution is $z=N-2a-4b-2$. Thus, for the term to be a modification of GR, the condition on $N$ must be $N \geq 2a + 4b$.  If such term exists in the Lagrangian then the Lagrangian is a modification of the Carrollian limit of GR. The $n$-th order is given by the modified optimization problem with $z \to z+2n$.

We can see that the conditions on $N$ in the two limits are mutually exclusive. Thus, no such term (of the form \eqref{HDG Lagrangian}) with finite $a$, $b$ can be a GR modification in the two limits simultaneously. This implies that no Lagrangian of the form $f(g_{\mu\nu}, R_{\mu\nu\sigma\rho})$ with polynomial $f$ can be a modification of GR in both limits simultaneously. In both limits we can recover the results from the previous section if we set $\mathcal{R}= \Bar{\mathcal{R}} = R$. We can recover quadratic gravity examples in the previous section by setting $a=2,b=0$ or $a=0,b=2$ or $a=b=1$.

\subsection{$f(g_{\mu\nu},R_{\mu\nu\sigma\rho},\nabla_{\mu})$ gravity}

Now we consider the most general HDG theory, i.e., $f(g_{\mu\nu},R_{\mu\nu\sigma\rho},\nabla_{\mu})$, where $f$ is polynomial. Following \cite{Biswas:2016egy}, any such theory can be recast to the form
\begin{equation}\label{general lagrangian}
    \mathcal{L} = \mathcal{P}_0 + \sum_i \mathcal{P}_i \prod_I O_{iI}\mathcal{Q}_{iI},
\end{equation}
where $\mathcal{P}_0$ and $\mathcal{Q}_{iI}$ are tensors made up of the Riemann tensor and the metric. The symbols $O_{iI}$ denote differential operators that are constructed from contractions of covariant derivatives. Notice that $\mathcal{P}_0$ is just $f(g_{\mu\nu},R_{\mu\nu\sigma\rho})$ which was discussed in the previous subsection. The second term of \eqref{general lagrangian} can be rearranged and written in our notation as
\begin{equation}
   \mathcal{L} = \alpha' c^N \sum_{i=1}^{i_{\textrm{max}}}\mathcal{R}^{a_i}\prod_{j=1}^{j_{\textrm{max}}}\nabla^{b_{ij}}\bar{\mathcal{R}}^{c_{ij}},
\end{equation}
where we added a constant with no $c$ dependency $\alpha'$. Moreover, here $i_{\textrm{max}}$, $j_{\textrm{max}}$, $a_i$, $b_{ij}$, $c_{ij}$ and $N$ are integers. Expanding the terms using \eqref{HDG expansion} we get
\begin{equation}
    \begin{aligned}
        \mathcal{R}^{a_i} &= \sum_{\mathclap{\substack{ \sum\limits_{n=0}^4k_{in}=a_i \\ k_{in}\geq 0}}}\frac{a_i!}{k_{i1}!...k_{in}!}\prod_{t=1}^{4}c^{2k_{it}(t-2)}(\mathcal{R}_t)^{k_{it}}, 
        \\
        \bar{\mathcal{R}}^{c_{ij}} &= \sum_{\mathclap{\substack{ \sum\limits_{m=0}^4\bar{k}_{ijm}=c_{ij} \\ \bar{k}_{ijm}\geq 0}}}\frac{c_{ij}!}{\bar{k}_{ij1}!...\bar{k}_{ijm}!}\prod_{\bar{t}=1}^{4}c^{2\bar{k}_{ij\bar{t}}(\bar{t}-3)}(\bar{\mathcal{R}}_{\bar{t}})^{\bar{k}_{ij\bar{t}}}, 
        \\
        \nabla^{b_{ij}} &= \sum_{\mathclap{\substack{\alpha_{ij},\beta_{ij},\gamma_{ij} \geq 0 \\ \alpha_{ij}+\beta_{ij}+\gamma_{ij}=b_{ij}}}}c^{-2\alpha_{ij}+2\gamma_{ij}}\frac{b_{ij}!}{\alpha_{ij}!\beta_{ij}!\gamma_{ij}!}\nabla_1^{\alpha_{ij}}\nabla_3^{\gamma_{ij}}\nabla_2^{\beta_{ij}}.
    \end{aligned}
\end{equation}
Putting everything together, we arrive at the action
\begin{equation}\label{general HDG action}
    \begin{aligned}
      \mathcal{L} &= \alpha' \sum_{i=1}^{i_{\textrm{max}}}\sum_{\substack{\sum\limits_{n=0}^{a_i}k_{in}=a_i \\ k_{in}\geq 0}}\prod_{j=1}^{j_{\textrm{max}}} \sum_{\substack{ \sum\limits_{m=0}^4\bar{k}_{ijm}=c_{ij} \\ \bar{k}_{ijm}\geq 0}} \sum_{\substack{\alpha_{ij},\beta_{ij},\gamma_{ij} \geq 0 \\ \alpha_{ij}+\beta_{ij}+\gamma_{ij}=b_{ij}}} \frac{a_i!}{k_{i1}!...k_{in}!}\frac{c_{ij}!}{\bar{k}_{ij1}!...\bar{k}_{ijm}!}\frac{b_{ij}!}{\alpha_{ij}!\beta_{ij}!\gamma_{ij}!} \\ 
      & \feq\times \nabla_1^{\alpha_{ij}}\nabla_3^{\gamma_{ij}} \prod_{t=1}^{4}\prod_{\bar{t}=1}^4 c^{2k_{it}(t-2)+2\bar{k}_{ij\bar{t}}(\bar{t}-3)-2\alpha_{ij}+2\gamma_{ij} + N} (\mathcal{R}_t)^{k_{it}} \nabla_2^{\beta_{ij}} (\mathcal{R}_{\bar{t}})^{\bar{k}_{ij\bar{t}}}.   
    \end{aligned}
\end{equation}
As before, we will now discuss how the Galilean and Carrollian expansions can be transformed into constrained optimization problems.

\subsubsection{Galilean limit}
The LO of the Galilean limit is given by $i_{\textrm{max}}$ maximization problems
\begin{equation}
    z_i = \max\bigg[\sum_{j=1}^{j_{\textrm{max}}}\bigg(2\sum_{t=1}^{4}k_{it}(t-2) + 2\sum_{\bar{t}=1}^4 \bar{k}_{ij\bar{t}}(\bar{t}-3) -2\alpha_{ij} + 2 \gamma_{ij} + N \bigg)\bigg],
\end{equation}
subjected to the constraints
\begin{equation}
\begin{gathered}
    \sum_{n=0}^4 k_{in} =a_i,  \hspace{10pt} \sum_{m=0}^4 \bar{k}_{ijm} = c_{ij},  \hspace{10pt} \alpha_{ij}+\beta_{ij}+\gamma_{ij}=b_{ij}, 
    \\
    k_{in},\bar{k}_{ijm},\alpha_{ij},\beta_{ij},\gamma_{ij} \geq 0.
\end{gathered}
\end{equation}
The $n$-th order in the Galilean expansion is given by similar optimization problems with $z_i \to z_i-2n$. All such problems are solvable using the simplex method.

\subsubsection{Carrollian limit}
The LO of the Carrollian expansion is given by a similar problem to the Galilean one but with minimization instead of maximization i.e,
\begin{equation}
    z_i = \min\bigg[\sum_{j=1}^{j_{\textrm{max}}}\bigg(2\sum_{t=1}^{4}k_{it}(t-2) + 2\sum_{\bar{t}=1}^4 \bar{k}_{ij\bar{t}}(\bar{t}-3) -2\alpha_{ij} + 2 \gamma_{ij} + N \bigg)\bigg],
\end{equation}
subjected to the constraints
\begin{equation}
\begin{gathered}
    \sum_{n=0}^4 k_{in} =a_i,  \hspace{10pt} \sum_{m=0}^4 \bar{k}_{ijm} = c_{ij},  \hspace{10pt} \alpha_{ij}+\beta_{ij}+\gamma_{ij}=b_{ij}, 
    \\
    k_{in},\bar{k}_{ijm},\alpha_{ij},\beta_{ij},\gamma_{ij} \geq 0.
\end{gathered}
\end{equation}
The $n$-th order is given by similar problem with $z_i \to z_i + 2n$.

For this type of terms to be modifications of GR in both limits in the LO (in the case of a theory with a Lagrangian of the form $R+\mathcal{L}$, where $\mathcal{L}$ is the Lagrangian in \eqref{general HDG action}), there must exist $i_1$ and $i_2$ such that
\begin{equation}\label{GR modification conditions}
    \begin{aligned}
   \max\bigg[\sum_{j=1}^{j_{\textrm{max}}}\bigg(2\sum_{t=1}^{4}k_{i_1t}(t-2) + 2\sum_{\bar{t}=1}^4 \bar{k}_{i_1j\bar{t}}(\bar{t}-3) -2\alpha_{i_1j} + 2 \gamma_{i_1j} \bigg)\bigg] &=2-N, \\
   \min\bigg[\sum_{j=1}^{j_{\textrm{max}}}\bigg(2\sum_{t=1}^{4}k_{i_2t}(t-2) + 2\sum_{\bar{t}=1}^4 \bar{k}_{i_2j\bar{t}}(\bar{t}-3) -2\alpha_{i_2j} + 2 \gamma_{i_2j}\bigg)\bigg] &= -N-2.  
    \end{aligned}
\end{equation}
If such terms exist, then the Lagrangian is a modification of GR in both limits simultaneously. We have shown that theories of the form \eqref{HDG Lagrangian} as well as quadratic gravity do not satisfy these conditions. We will leave the search for such theory (if any exists) to future work. 


\section{Conclusions}\label{sec7}
In this paper, we reviewed the methods used to construct non-Lorentzian gravitational theories from Lorentzian ones. We showed that all methods of taking the non-Lorentzian limits lead to the same metric, thus, the same non-Lorentzian theories. In the case of the Galilean expansion, the GAG procedure gives the same Galilean theories as the PNR approach. However, the PNR approach explore more theories from expanding the PNR quantities in powers of $c^{-2}$. Such expansion gives rise to infinite number of theories, one for each order, which are not accounted for in other methods. However, these theories are not Galilean till the NLO. The same Galilean theories can be deduced by performing the ADM decomposition and taking the infinite limit of the norm of the orthogonal vector, the IS approach. A similar situation occurs in the Carrollain expansion, although the ZS approach where the limit of the norm of the orthogonal vector is sent to zero (which give the same Carrollian theories as the Carroll group gauging approach) is computationally easier than the PUL approach. The PUL approach explores a larger space of theories, the Carrollian theories to the NLO coincide with the theories we get from other approaches by truncating the expansions of the PUL quantities setting all higher order fields to zero.

Having established that all approaches give the same metric, one can write general expansions for the curvature tensors. We explored the most general HDG theory with polynomial $f$, introducing an algorithm to calculate the $n$-th order of its Galilean and Carrollian expansions as follows:
\begin{enumerate}
    \item We write the Lagrangian and identify which form it takes (is it of the form \eqref{f(R) Lagrangian}, \eqref{HDG Lagrangian}, or the general \eqref{general HDG action}) and if the Lagrangian depends on $c$ explicitly.
    \item We identify the parameters in the respective section of this paper by comparing the given Lagrangian with the expanded one.
    \item We deduce the optimization problem equivalent to the Lagrangian's Galilean or Carrollian expansion by substituting the parameters from the previous step into the respective optimization problem in the relevant section.
    \item We solve the optimization problem (manually or using a computer) to get the desired order of the expansion.
\end{enumerate}

Higher orders in the Galilean expansion are useful to get more accurate results of dynamical systems in the post Newtonian approximation. Thus, by transforming the problem into computationally easier optimization problems, we can study such systems more efficiently. We leave the analysis of such dynamical systems to future work. On the other hand, although higher orders in the Carrollian expansion have no utility at the moment, some may be discovered in the near future given the increasing interest in Carrollian physics, and Carrollian gravity in particular, and having an algorithm to compute such higher order will be beneficial then.

Another interesting future direction of research is to search for gravity theories which are viable modifications of GR in both Carrollian, and Galilean limits at the same time i.e, satisfies the conditions \eqref{GR modification conditions}. If such a theory exists, it is interesting to see what significance it has. Is there a defining property of the theory that allows this? If so, how does it impact its solutions. It is also interesting to apply the algorithm introduced to study its Galilean and Carrollian limits, and see how the GR black holes get modified.


\section*{Acknowledgements}

The authors would like to thank Eric Bergshoeff (Groningen, Netherlands), Marc Henneaux (Brussels, Belgium), Niels Obers (Nordita, Sweden) and Rodrigo Olea (Santiago, Chile) for stimulating discussions. P.T. and I.K. were supported by Primus grant PRIMUS/23/SCI/005 from Charles University.


\appendix
\section{List of formulae}
In this appendix we list the formulae for the terms of the expansion of curvature tensors used in the paper. The following formulae are written in the PUL quantities. They are derived by direct computations from the metric. Notice that the quantities presented are not expanded in powers of $c$, and will appear in the $n-$th order Lagrangian.
\begin{itemize}
    \item Riemann tensor with all indices down $R_{\sigma \lambda\mu\nu}$ :
    \begin{equation}
    \begin{aligned}
        \mathcal{R}_1 &= 2\mathcal{K}_{\lambda[\nu}\mathcal{K}_{\mu]\sigma},\\
        \mathcal{R}_2 &= \overset{c}{R}_{\sigma \lambda\mu\nu} +  2T_{\sigma}\nabla_{[\nu}\mathcal{K}_{\mu]\lambda} + \mathcal{K}_{\lambda\alpha}C^{\alpha}_{[\nu\mu]}T_{\sigma} + 2T_{\sigma}T_{\lambda}\mathcal{K}^{\alpha}_{[\nu}\mathcal{K}_{\mu]\alpha}+ 2\nabla_{[\mu}(\mathcal{K}_{\nu]\sigma}T_{\lambda})\\ & \feq+ 2T_{\lambda}C^{\alpha}_{[\mu\nu]}\mathcal{K}_{\alpha\sigma}+T_{(\mu}B_{\alpha)\sigma}V^{\alpha}\mathcal{K}_{\nu\lambda} - T_{(\nu}B_{\alpha)\sigma}V^{\alpha}\mathcal{K}_{\mu\lambda},\\
        \mathcal{R}_3 &= T_{\sigma}\mathcal{K}_{\mu\alpha}T_{(\nu}B_{\lambda)}{}^{\alpha} - T_{\sigma}\mathcal{K}_{\nu\alpha}T_{(\mu}B_{\lambda)}{}^{\alpha} + \nabla_{\nu}(T_{(\mu}B_{\lambda)\sigma}) - \nabla_{\mu}(T_{(\nu}B_{\lambda)\sigma})\\ & \feq+ 2C^{\alpha}_{[\nu\mu]}T_{(\alpha}B_{\lambda)\sigma} - T_{(\mu}B_{\alpha)\sigma}T_{\lambda}\mathcal{K}_{\nu}^{\alpha}  + T_{(\nu}B_{\alpha)\sigma}T_{\lambda}\mathcal{K}_{\mu}^{\alpha},\\
        \mathcal{R}_4 &= T_{(\mu}B_{\sigma)\alpha}T_{(\nu}B_{\lambda)}{}^{\alpha}.
        \end{aligned}
    \end{equation} 
    where $B_{\mu\nu}= \partial_{\mu}T_{\nu}-\partial_{\nu}T_{\mu}$, $C^{\rho}_{\mu\nu}$ is the connection in \eqref{PUL connection}, $\nabla_{\mu}$ is its compatible covariant derivative, and $ \overset{c}{R}_{\sigma \lambda\mu\nu}$ is its Riemann tensor.
    \item Riemann tensor with all indices up $R^{\rho \alpha \beta \gamma}$ :
    \begin{equation}
    \begin{aligned}
        \Bar{\mathcal{R}}_1 &= -V^{\alpha}V^{\beta}V^{\mu}\Pi^{\gamma\nu}\nabla_{\mu}\mathcal{K}_{\nu}^{\rho} - 2 V^{\alpha}V^{\beta}V^{\mu}\Pi^{\gamma\nu}C^{\sigma}_{[\mu\nu]}\mathcal{K}_{\sigma}^{\rho}-\tfrac{1}{2}V^{\alpha}V^{\beta}V^{\rho}V^{\lambda}\mathcal{K}_{\sigma}^{\gamma}B_{\lambda}{}^{\sigma} - V^{\alpha}V^{\gamma}V^{\nu}\Pi^{\mu\beta}\nabla_{\nu}\mathcal{K}_{\mu}^{\rho} \\ &
        \feq- 2V^{\alpha}V^{\gamma} V^{\nu}\Pi^{\mu\beta}C^{\sigma}_{[\mu\nu]}\mathcal{K}_{\sigma}^{\rho} - V^{\alpha}V^{\gamma}V^{\rho}\mathcal{K}_{\sigma}^{\beta}B_{\nu}{}^{\sigma} + 2V^{\alpha}V^{\rho}\mathcal{K}^{\alpha[\gamma}\mathcal{K}_{\alpha}^{\beta]} + V^{\beta}V^{\rho}\Pi^{\nu\gamma}\Pi^{\alpha\lambda}\nabla_{\mu}\mathcal{K}_{\nu\lambda}\\ &
        \feq-2 V^{\beta}V^{\rho}V^{\mu}\mathcal{K}_{\sigma}^{\alpha}\Pi^{\nu\gamma}C^{\sigma}_{[\nu\mu]} - 2V^{\gamma}V^{\nu}V^{\rho}\Pi^{\mu\beta}\Pi^{\alpha\lambda}\nabla_{[\nu}\mathcal{K}_{\mu]\lambda} - 2V^{\gamma}V^{\nu}V^{\rho}\Pi^{\mu\beta}\Pi^{\alpha\lambda}\mathcal{K}_{\lambda\sigma}C^{\sigma}_{[\nu\mu]}\\
        \Bar{\mathcal{R}}_2 &= V^{\alpha}V^{\beta}V^{\gamma}V^{\nu}B^{\sigma\rho}B_{\sigma\nu} + \tfrac{1}{2}V^{\alpha}V^{\beta}V^{\mu}\Pi^{\nu\gamma}\nabla_{\nu}(T_{\lambda}B_{\mu}{}^{\rho}) - V^{\alpha}V^{\beta}V^{\mu}V^{\lambda}\Pi^{\nu\gamma}\nabla_{\mu}(T_{(\nu}B_{\lambda)}{}^{\rho})\\ & 
        \feq-V^{\alpha}V^{\beta}V^{\mu}\Pi^{\nu\gamma}B_{\sigma}{}^{\rho}C^{\sigma}_{[\nu\mu]} + \tfrac{1}{2}V^{\alpha}V^{\beta}B_{\sigma}{}^{\rho}\mathcal{K}^{\gamma\sigma} + V^{\alpha}V^{\gamma}V^{\lambda}V^{\nu}\Pi^{\mu\beta}\nabla_{\nu}(T_{(\mu}B_{\lambda)}{}^{\rho}) \\ &
        \feq-V^{\alpha}V^{\beta}V^{\lambda}V^{\nu} \Pi^{\mu\beta}\nabla_{\mu}(T_{(\nu}B_{\lambda)}{}^{\rho}) - V^{\alpha}V^{\beta}V^{\nu}\Pi^{\mu\beta}B_{\sigma}{}^{\rho}C^{\sigma}_{[\nu\mu]}- V^{\alpha}V^{\beta}V^{\nu}\mathcal{K}^{\beta\sigma}T_{(\nu}B_{\sigma)}{}^{\rho}\\ &
        \feq+ 2 V^{\alpha}\Pi^{\mu\beta}\Pi^{\nu\gamma}\nabla_{[\mu}\mathcal{K}_{\nu]}^{\rho} + V^{\alpha}V^{\rho}\mathcal{K}_{\sigma}^{[\gamma}B^{\beta]\sigma} +  V^{\beta}V^{\gamma}V^{\mu}V^{\nu}\Pi^{\lambda\alpha}\nabla_{\nu}(T_{(\mu}B_{\lambda)}^{\rho})+2\mathcal{K}^{\alpha[\gamma}\mathcal{K}^{\beta]\rho} \\ & 
        \feq-  V^{\beta}V^{\gamma}V^{\mu}V^{\nu}\Pi^{\lambda\alpha}\nabla_{\mu}(T_{(\nu}B_{\lambda)}{}^{\rho})- 2V^{\beta}V^{\mu}\Pi^{\nu\gamma}\Pi^{\alpha\lambda}\nabla_{[\mu}(\mathcal{K}_{\nu]}^{\rho}T_{\lambda}) - \tfrac{1}{2}V^{\beta}\mathcal{K}_{\sigma}^{\gamma}B^{\alpha\sigma}-V^{\beta}V^{\mu}\mathcal{K}^{\alpha\gamma}B_{\mu}{}^{\rho} \\ &
        \feq- 2V^{\gamma}V^{\nu}\Pi^{\mu\beta}\Pi^{\alpha\lambda}\nabla_{[\mu}(\mathcal{K}_{\nu]}^{\rho}T_{\lambda}) + \tfrac{1}{2}V^{\gamma}V^{\rho}\mathcal{K}_{\sigma}^{\beta}B^{\alpha \sigma} + V^{\gamma}V^{\sigma}B_{\sigma}{}^{\rho}\mathcal{K}^{\alpha\beta} + 2V^{\rho}\Pi^{\alpha\lambda}\Pi^{\beta\mu}\Pi^{\nu\gamma}\nabla_{[\nu}\mathcal{K}_{\mu]\lambda}
        \\
        \Bar{\mathcal{R}}_3 &=\overset{c}{R}{}^{\rho\alpha\beta\gamma}+ \tfrac{1}{4}V^{\alpha}V^{\beta}B_{\sigma}{}^{\rho}B^{\sigma \rho}-V^{\alpha}V^{\lambda}\Pi^{\mu\beta}\Pi^{\nu\gamma}\nabla_{\nu}(T_{(\nu}B_{\lambda)}{}^{\rho}) + V^{\alpha}V^{\lambda}\Pi^{\mu\beta}\Pi^{\nu\gamma}\nabla_{[\mu}\mathcal{K}_{\nu]}^{\rho}+\tfrac{1}{4}V^{\beta}V^{\gamma}B_{\sigma}{}^{\rho}B^{\alpha\sigma} \\&
        \feq-V^{\beta}V^{\mu}\Pi^{\nu\gamma}\Pi^{\alpha\lambda}\nabla_{\nu}(T_{(\mu}B_{\lambda)}{}^{\rho}) + V^{\beta}V^{\mu}\Pi^{\gamma \nu}\Pi^{\alpha \lambda}\nabla_{\mu}(T_{(\nu}B_{\lambda)}{}^{\rho}) -V^{\gamma}V^{\nu}\Pi^{\mu\beta}\Pi^{\alpha\lambda}\nabla_{\nu}(T_{(\mu}B_{\lambda)}{}^{\rho})\\ &
        \feq+ V^{\gamma}V^{\nu}\Pi^{\mu\beta}\Pi^{\alpha\lambda}\nabla_{\mu}(T_{(\nu}B_{\lambda)}{}^{\rho}) + 2\Pi^{\alpha\lambda}\Pi^{\mu\beta}\Pi^{\nu\gamma}\nabla_{[\mu}(\mathcal{K}_{\nu]}^{\rho}T_{\lambda})-\tfrac{1}{2}B^{\beta\rho}\mathcal{K}^{\alpha\gamma}+\tfrac{1}{2}B^{\gamma\rho}\mathcal{K}^{\alpha\beta}\\
        \Bar{\mathcal{R}}_4 &= \Pi^{\alpha\lambda}\Pi^{\mu\beta}\Pi^{\nu\gamma}\nabla_{\nu}(T_{(\mu}B_{\lambda)}{}^{\rho})-\Pi^{\alpha\lambda}\Pi^{\mu\beta}\Pi^{\nu\gamma}\nabla_{\mu}(T_{(\nu}B_{\lambda)}{}^{\rho})\\
        \end{aligned}
    \end{equation}
    \item Ricci tensor with indices down $R_{\mu\nu}$
    \begin{equation}
        \begin{aligned}
        \mathcal{R}_1 &= -\nabla_{\sigma}(V^{\sigma}\mathcal{K}_{\mu\nu}) - 2V^{\sigma}C^{\rho}_{[\sigma \mu]}\mathcal{K}_{\nu\rho} + \mathcal{K}_{\mu\nu}\mathcal{K}-\mathcal{K}_{\mu\sigma}\mathcal{K}_{\nu}^{\sigma}\\
        \mathcal{R}_2 &= \overset{c}{R}_{\mu\nu}+ \nabla_{\sigma}(T_{\nu}\mathcal{K}_{\mu}^{\sigma})-\nabla_{\nu}(T_{\mu}\mathcal{K}) + 2C^{\rho}_{[\nu\sigma]}T_{\mu}\mathcal{K}_{\rho}^{\sigma} + \mathcal{K}_{(\mu}^{\alpha}B_{\nu)\alpha} - V^{\sigma}\mathcal{K}_{(\nu}^{\alpha}T_{\mu)}B_{\sigma \alpha} \\
        \mathcal{R}_3 &= -\nabla_{\sigma}(T_{(\nu}B_{\mu)}{}^{\sigma}) + 2C^{\sigma}_{[\nu\rho]}T_{(\sigma}B_{\mu)}{}^{\rho} + T_{(\nu}B_{\sigma)}{}^{\rho}T_{\mu}\mathcal{K}_{\rho}^{\sigma}\\
        \mathcal{R}_4 &= -\tfrac{1}{4}T_{\mu}T_{\nu}B_{\alpha\beta}B^{\alpha\beta}\\
        \end{aligned}
    \end{equation}
    \item Ricci tensor with indices up $R^{\sigma\rho}$
    \begin{equation}
        \begin{aligned}
        \Bar{\mathcal{R}}_1 &= V^{\sigma}V^{\rho}V^{\nu}\nabla_{\nu}\mathcal{K} - 2V^{\sigma}V^{\rho}V^{\nu}C^{\alpha}_{[\nu\beta]}\mathcal{K}_{\alpha}^{\beta}\\
        \Bar{\mathcal{R}}_2 &= -V^{\sigma}V^{\rho}V^{\mu}\nabla_{\alpha}B_{\mu}{}^{\alpha} - V^{\rho}\Pi^{\mu\sigma}\nabla_{\alpha} \mathcal{K}_{\mu}^{\alpha} V^{\rho}V^{\nu}\Pi^{\nu\sigma}\nabla_{\nu}(T_{\mu}\mathcal{K}) -V^{\rho}V^{\nu}\mathcal{K}^{\alpha\sigma}B_{\nu\alpha} + V^{\sigma}\Pi^{\nu\rho}\nabla_{\nu}\mathcal{K} \\ &
        \feq- \nabla_{\alpha}(V^{\alpha}\mathcal{K}^{\sigma\rho}) - 2V^{\alpha}C^{\beta}_{[\alpha \mu]}\mathcal{K}^{\rho}_{\beta}\Pi^{\mu\sigma} + \mathcal{K}\mathcal{K}^{\sigma\rho} - \mathcal{K}^{\alpha\sigma}\mathcal{K}^{\rho}_{\alpha}\\
        \Bar{\mathcal{R}}_3 &= \overset{c}{R}{}^{\sigma\rho}-\tfrac{1}{4}V^{\sigma}V^{\rho}B_{\alpha\beta}B^{\alpha\beta} + V^{\rho}V^{\nu} \Pi^{\mu\sigma}\nabla_{\alpha}(T_{(\nu}B_{\mu)}{}^{\alpha}) + \Pi^{\nu\rho}V^{\sigma}V^{\mu}\nabla_{\alpha}(T_{(\nu}B_{\mu)}{}^{\alpha}) \\ &
        \feq+ \Pi^{\mu\sigma}\Pi^{\nu\rho}\nabla_{\alpha}(T_{\nu}\mathcal{K}_{\mu}^{\alpha}) -  \Pi^{\mu\sigma}\Pi^{\nu\rho}\nabla_{\nu}(T_{\mu}\mathcal{K}) + \Pi^{\mu\sigma}\Pi^{\nu\rho}\mathcal{K}^{\alpha}_{(\mu}B_{\nu)\alpha}\\
        \Bar{\mathcal{R}}_4 &= -\Pi^{\nu\rho}\Pi^{\mu\sigma}\nabla_{\alpha}(T_{(\nu}B_{\mu)}{}^{\alpha})\\
        \end{aligned}
    \end{equation}
    \item Ricci scalar $R$
    \begin{equation}\label{ricci scalar}
        \begin{aligned}
          \mathcal{R}_1 =  \Bar{\mathcal{R}}_1 &= \mathcal{K}^{2}-\mathcal{K}_{\mu\nu}\mathcal{K}^{\mu\nu} - 2\nabla_{\nu}(V^{\nu}\mathcal{K})\\
       \mathcal{R}_2 = \Bar{\mathcal{R}}_2 &= -\overset{c}{R} - \nabla_{\mu}(V^{\nu}B_{\nu}{}^{\mu})\\
       \mathcal{R}_3 = \Bar{\mathcal{R}}_3 &= \tfrac{1}{4}B_{\mu\nu}B^{\mu\nu}\\
        \end{aligned}
    \end{equation}
    \item The covariant derivative $\nabla_{\mu}$
    \begin{equation}
        \begin{aligned}
            (\nabla_1)_{\mu}N_{\nu} &= -V^{\rho}\mathcal{K}_{\mu\nu}N_{\rho}\\
            (\nabla_2)_{\mu}N_{\nu} &= \partial_{\mu}N_{\nu} + \Pi^{\rho\lambda}T_{\nu}\mathcal{K}_{\mu\lambda}N_{\rho} \\
            (\nabla_3)_{\mu}N_{\nu} &= - T_{(\mu}B_{\nu)\lambda}\Pi^{\rho\lambda}N_{\rho} \\
        \end{aligned}
    \end{equation}
    where $\partial_{\mu}$ is the partial derivative with respect to a generic coordinate system.
\end{itemize}

\end{document}